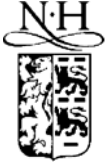

# Spectroscopy of ultralight bosons by photon-boson-photon conversion-reconversion: experimental aspects

*Ugo Gastaldi**

*INFN-Laboratori Nazionali di Legnaro*

**Viale dell'Università 2, I-35020 Legnaro (Pd), Italy**

**Abstract**

A conversion-reconversion experimental scheme is proposed where photons stored in a Fabry-Perot cavity (FP) embracing a dipole conversion magnet generate two beams of ultralight bosons emitted in opposite directions. Fractions of the bosons convert in dipole magnets aligned along the two boson beam lines and generate photons of the same energy and directions of the FP grandparent photons. The reconversion photons are detected individually by means of single photon detectors. Such an experiment can determine the bosons parity by measurements with different angles between the polarization vector of linearly polarized light injected into the FP and the direction of the magnetic field $B_c$ of the conversion magnets. It can establish both the bosons mass m and coupling constant to two photons $g_{m\gamma\gamma}$ by measurements with different lengths $L_R$ of the string of powered regeneration magnets of one boson beam line, while the second boson beam line is left unchanged and used for normalisation. The optimal total length $L_C$ of the string of conversion magnets depends on $m^2$ and on the energy E of the laser photons. For E = 2.4 eV $L_C$ is of the order of 3 m if m = $10^{-3}$ eV and of the order of 300 m if m=$10^{-4}$ eV. If the signals observed by the PVLAS experiment are due to the existence of ultralight bosons with m = $10^{-3}$ eV and $g_{m\gamma\gamma}$ = $10^{-6}$ GeV$^{-1}$, these regeneration measurements are feasible with existing technologies on a site of length below 100 m in relatively short periods of data taking time.

## 1. Introduction

Several suggestions to ascertain the existence of ultralight or zero mass spin zero bosons exploit, for bosons with non zero coupling to two photons, their interactions with electromagnetic fields [1-6]. A detector on earth could be immersed in a sea of nonrelativistic ultralight bosons (cold dark stuff), or be traversed by relativistic ultralight bosons emitted by the sun, or bombarded by a beam of relativistic ultralight bosons generated in the lab by coherent Primakoff conversion [7] of laser photons interacting

---
*Tel.+39-049-8068380; fax 049-641925; email: gastaldi@lnl.infn.it

with the virtual photons of a static magnetic or electric field.

A pilot regeneration experiment was done at BNL by the BFST Collaboration and produced upper limits for the coupling of ultralight bosons to two photons[8,9]. Ambitious scaled-up versions of the BFST set-up have been recently suggested [10,11]. Here we propose a new experimental arrangement (shown schematically in fig.1), which is a generalization of the BNL set-up, adopts the optical techniques developed in the PVLAS experiment [12] at LNL and introduces a normalization arm, that corresponds to the near detector in long base line neutrino oscillation experiments with neutrino beams (from acclerators or reactors) directed towards the far (main) detector [13].

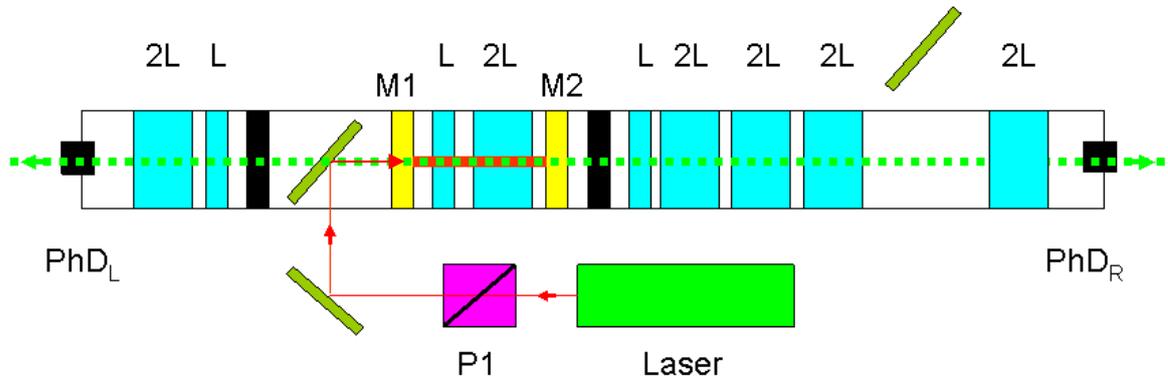

Fig.1 Schematic lay-out of the regeneration set-up. M1, M2: mirrors of the Fabry-Perot cavity. L, 2L: dipole magnets of length L an 2L with vertical magnetic field B direction; the polarity of the magnetic field can be chosen separately for each magnet. P1: polarizer; it can be rotated in order to change the angle between the polarization of the light stored in the Fabry-Perot and the magnetic field of the dipoles. $PhD_L$ and $PhD_R$: single photon detectors.

Two dipole magnets of field intensity $B_c$ and length L and 2L are positioned between the mirrors M1 and M2 of a Fabry–Perot cavity (FP) and are aligned with the FP light path. A laser followed by a polarizer P1 injects light into the FP cavity. The polarizer P1 selects photons of the laser beam with polarization (direction of the electric field of the photons) directed e.g. vertically. The laser is locked to the FP cavity by the same techniques [14,15] used in PVLAS. In a FP of finesse F photons travel back and forth N/2 times with $N=(2/\pi)F$ since they on average undergo a total of N reflections between the mirrors M1 and M2. If ultralight spin zero bosons exist in nature and have a non zero coupling $g_{m\gamma\gamma}$ to two photons, two beams of bosons produced by Primakoff conversion in the magnetized regions of the FP cavity emerge out of the FP cavity in two opposite directions.

The two boson beams have equal intensities

$$I_b = \frac{N}{2}\frac{W}{E}C \qquad (1)$$

W is the power of the polarized laser beam that traverses the FP cavity, E is the energy of the laser photons, C is the probability of photon to boson conversion in the conversion dipole magnets.

The two boson beams traverse undisturbed the FP mirrors and all the materials that decouple the conversion from the regeneration regions (e.g. mirrors for feeding and monitoring the FP cavity, monitors of light reflected and transmitted by the FP, optical shutters that prevent photons lost by the FP to enter the regeneration lines), since these bosons interact with matter much more weakly than neutrinos do.

Two strings or reconversion dipole magnets are positioned along the trajectories of the two boson beams. The left string has two dipole magnets of length L and 2L aligned on the left of the conversion region and is used for normalization. The right string features several identical dipole magnets of length 2L after a shorter one of length L and is used for measurements with different lengths of the magnetized region, in analogy to an idealized long base line neutrino experiment where the length of the baseline could be changed in well known discrete steps. Each of the magnets is powered independently, and is normally off, or at maximum field with positive vertical polarity. It is also possible to invert the polarity of any magnet. By these means the effective total magnetized lengths $L_C$ of the conversion region, $L_L$ of the left regeneration region and $L_R$ of the right regeneration region can be varied respectively in the windows

$L \leq L_C \leq 3L$
$L \leq L_L \leq 3L$
$L \leq L_R \leq L + 2nL$

where n is the number of magnets of length 2L powered with the same polarity.

Ultralight bosons can reconvert into photons of the same energy and direction of the laser photons by inverse Primakoff effect in the magnets of the two boson beam lines as a result of interactions with virtual photons of the magnetic field.

The rates $Ph_L$ and $Ph_R$ of photon production in the left and right boson beams are respectively:

$$Ph_{L/R} = I_b R_{L/R} = \frac{N}{2} \frac{W}{E} C R_{L/R} \qquad (2)$$

$R_L$ and $R_R$ are respectively the probability of boson to photon conversion in the left and right string of magnets. We shall see in the following that these probabilities are controlled by the effective total magnetized lengths $L_L$ and $L_R$ of the left and right strings of magnets.

Since the regeneration photons are all emitted along the two opposite boson beam directions, they can all be detected with the same geometrical efficiency by two photon detectors $PhD_L$ and $PhD_R$ positioned at the end of each regeneration line.

The signal rates $S_L$ and $S_R$ in the two photon detectors will be

$$S_{L/R} = \eta_{L/R} Ph_{L/R} = \eta_{L/R} \frac{N}{2} \frac{W}{E} C R_{L/R} \qquad (3)$$

$\eta_L$ and $\eta_R$ are the detection efficiencies.

The ratio

$$\frac{S_R}{S_L} = \frac{\eta_R R_R}{\eta_L R_L} \qquad (4)$$

factorizes the laser power W and the FP finesse F and their variations or fluctuations.

Measurements with different orientations of the input polarimeter P1 will give the experimental inputs to determine a) the existence of ultralight bosons, b) if they are only of one parity or if species of both parities exist, and c) the ultralight boson parity, in the case only one species exists.

If only bosons of one parity exist, measurements with different settings of the string of right magnets give the experimental inputs to determine the mass m and the coupling constant to two photons $g_{m\gamma\gamma}$.

If our universe features the existence of several ultralight boson species, several sets of measurements performed with different energies of the laser beam would be necessary to clarify the scenario.

## 2. Photon and boson beams

For this kind of experiments it is necessary to maximize the total number of photons present in the FP cavity at any time. This corresponds to maximize the intensity of the photon beam injected into the FP cavity times the average number N of reflections of each photon stored in the cavity. W/E photons traverse the cavity per unit time. At any time there are then (W/E)(N/2) photons traversing the conversion magnet in each of the two opposite directions.

Let us assume the following parameters to get a feeling of the orders of magnitude: W = 0.24 Watt light power traversing the FP, E= 2.4 eV energy of the laser photons, N=$10^5$ average number of reflections in the FP. Since 1Watt = 6.25 $10^{18}$eV s$^{-1}$, W/E = 6.25 $10^{17}$ photons sec$^{-1}$ transit through the FP cavity and there are W/E N/2 = 3 $10^{22}$ photon traversals of the conversion region per second.

The photon to boson conversion probability C is given, for m<<E, by [16,9]

$$C = 4 g_{m\gamma\gamma}^2 B_C^2 E^2 m^{-4} \sin^2 \frac{m^2 L_C}{4E} \qquad (5)$$

The parameters m and $g_{m\gamma\gamma}$ are fixed by nature. We can change $B_C$, E and $L_C$. By increasing $B_C$, C grows proportionally to $B_C^2$. For fixed E, by increasing $L_C$ the term $m^2 L_C/4E$ grows and so does the term $\sin^2(m^2 L_C/4E)$ until $m^2 L_C/4E = \pi/2$.

There is no point in making the conversion magnet longher than $L_C = 2\pi\, E/m^2$, since the term $\sin^2$ cannot be greater than 1. For E = 2.4 eV and m=$10^{-3}$eV , the optimal conversion magnet length $L_C$ is about 3m, while for m=$10^{-5}$eV , the optimal conversion magnet length $L_C$ is about 30 Km.

In a scenario with $B_C$ =2 T, E= 2.4 eV and m = $10^{-3}$ eV, $g_{m\gamma\gamma}$ =1/ M = $10^{-6}$GeV$^{-1}$, no matter how long one could make $L_c$, it will always be

$$C < 4 g_{m\gamma\gamma}^2 B_C^2 E^2 m^{-4} < 4 \cdot 10^{-12} \quad (6)$$

If $L_C \ll 2\pi\, E/m^2$ , the conversion probability C can be approximated by C ≈ 1/4 $g_{m\gamma\gamma}^2$ $B_C^2$ $L_C^2$ , which does no more depend on m.

The reconversion probabilities $R_L$ and $R_R$ are given by expressions of the same type of the conversion probability:

$$R_{L/R} = 4 g_{m\gamma\gamma}^2 B_R^2 E^2 m^{-4} \sin^2 \frac{m^2 L_{L/R}}{4E} \quad (7)$$

Assuming that all the magnetic fields point in the same direction, e.g. upwards in the vertical direction, and an unobstructed optical path along each magnet string, with the parameters of the previous scenario (namely m = $10^{-3}$ eV, M = $10^6$GeV, E=2.4eV and B =2 T), the reconversion probabilities will be limited as indicated below, no matter how long the magnetized regions $L_L$ and $L_R$ of the left and right magnet strings could be made

$$R_{L/R} < 4 \cdot 10^{-12} \quad (8)$$

For $L_L \ll 2\pi\, E/m^2$ and $L_R \ll 2\pi\, E/m^2$, the reconversion probabilities can be approximated by the expressions $R_{L/R}$ ≈ 1/4 $g_{m\gamma\gamma}^2$ $B^2$ $L_{L/R}^2$, which do no more depend on m. We shall come back to this point in section 3.

The intensities of the two photon beams created by regeneration are given by

$$Ph_{L/R} = 8 N \frac{W}{E} g_{m\gamma\gamma}^4 B^4 E^4 m^{-8} \sin^2 \frac{m^2 L_C}{4E} \sin^2 \frac{m^2 L_{L/R}}{4E} \quad (9)$$

Assuming the physics parameters m = $10^{-3}$ eV, M = $10^6$GeV, the experimental parameters E=2.4 eV, W=0.24 Watt, N=$10^5$ for the laser and the FP cavity, and $B_C$= $B_L$ = $B_R$ =2T for the magnets, for any choice of $L_L$ and $L_R$ the rate of regenerated photons $Ph_{L/R}$ will not exceed 0.5phs sec$^{-1}$ .

In the experiment of ref. [8,9] single photons were detected by a photomultiplier 9893B/350 Thorn EMI cooled to -23$^0$C with detection efficiency η=0.1, a dark current rate in the single photoelectron peak $d_c$=0.6 Hz and a sensitive area for photons with a diameter Φ=9mm.

These performance figures can be used for conservative estimates of the sensitivities of the measurements that will be discussed later. It is worth noticing the large value of the diameter of the active area of the detector, which exceeds the typical diameter of the light beam envelope in the FP, allows to collect all the photons emitted along a boson beam line, no matter their production point, and protects the measurements against fluctuations of the FP beam alignment.

Use of TES cryogenic detectors can improve dramatically the detection and noise performances, since the single photon detection efficiency could grow to η=40% and the dark current rate could lower to $d_c$=$10^{-3}$ Hz in the single photoelectron peak [17], but the active area of these high performance devices would be about $10^{-2}$mm$^2$ . This last feature could render critical the alignment of the photon detectors at the extremities of both regeneration beam lines, because the axis of the FP should be defined with an accuracy and a stability of the order of 10 μm, in order to focus all photons produced in the two beam lines onto the sensitive parts of the two photodetectors. Focussing looks however possible because the photons are emitted along directions nearly parallel to the FP axis with maximum divergence of the order of the ratio r=$10^{-3}$ between the diameter of the FP light beam envelope and the distance between the two FP mirrors.

### 3. Physics

Axions are ultralight pseudoscalar bosons whose existence is requested to fix basic problems of QCD and could provide contribution to dark matter (for recent reviews see refs.[18-20]). The existence of ultralight scalar bosons (dilatons) could provide a basic experimental input for general relativity and cosmology (for reviews see refs.[21,22]) relevant for a particle description of dark matter and may be for the cosmological constant and the accelerated expansion of our universe.

The coupling constant to two gammas for an ultralight boson is not expected to be larger than $10^{-10}$ GeV$^{-1}$ for m=$10^{-3}$ eV both for pseudoscalars and scalars [23,24,18-20]. Therefore the indication of a coupling of the order of $10^{-6}$ GeV$^{-1}$ and a mass m=$10^{-3}$ eV emerging from the PVLAS data [25-27] is very surprising and requires confirmation. The surprise is twofold first because it suggests new physics at the mass scale around $10^6$ GeV, second because the visibility of the effects of the existence of ultralight bosons is enormously enhanced by the magnitude of the coupling. Several authors [28-31] have started exploring ways to interpret or cross check the PVLAS results and the relation with the limit $g_{m\gamma\gamma} < 1.16 \ 10^{-10}$GeV$^{-1}$ for m $< 2 \ 10^{-2}$eV established by the CAST experiment [32].

The PVLAS result is so surprising that, if confirmed, one should check experimentally whether the existence of only one boson of given parity can fit the data or two or more bosons of different parities are necessary. It is in this sense that we use the word spectroscopy. The large (of course comparatively to expectations) value of the coupling constant to two photons would enable this experimental effort in reasonable time.

Pseudoscalar bosons could be produced in the conversion magnet if the polarization vector of the light stored in the FP had a component parallel to the direction of the magnetic field $B_C$, since the interaction term of the effective lagrangian for a pseudoscalar coupling would be of the type $g_{m\gamma\gamma} \Phi \mathbf{E}_{photon} \cdot \mathbf{B}_C$ [1-5,9,16].

Scalar bosons could instead be produced in the conversion magnet if the polarization vector of the light stored in the FP had a component normal to $B_C$, since the interaction term would be of the type $g_{m\gamma\gamma} \Phi (\mathbf{E}_{photon} \times \mathbf{B}_C)$ [1-5,9,16].

If $\theta$ is the angle between the light polarization vector and the magnetic field vector $B_C$, the $\theta$ dependence for production of pseudoscalars is expected to be proportional to $\cos\theta$, while the production of scalars would be proportional to $\sin\theta$.

In the reconvertion magnets the direction of the magnetic field determines the polarization of the photons produced by inverse Primakoff effect. If the photon detector is not sensible to the polarization of the photons, the direction of the magnetic field of the regeneration magnets does not influence the detection rate. This fact is natural because for bosons with spin zero, all directions orthogonal to their motion are equivalent, and so the orientation of $\mathbf{B}_R$ could influence the polarization, but not the production rate of regeneration photons. By starting initial measurements with $\theta=45^0$, and powering the right string of magnets so to have the same lengths $L_R=L_L$ of magnets powered with the same field intensity B both in the left and right string of regeneration magnets, one expects $S_R/S_L=\eta_R/\eta_L$ $R_R/R_L = \eta_R/\eta_L$. The terms $S_R$ and $S_L$ on the left side of the above expression represent the differences of count rate between runs of equal duration with conversion magnets powered and not powered. If ultralight bosons are really produced and the normalization photon detector exibits a significant signal to noise ratio, the photon detector of the right string of magnets should have an equal number of counts (within errors) when $L_R=L_L$. If these circumstances would occur, at least one boson of yet unknown parity would be proven to exist.

By rotating the polarizer at the entrance of the FP, and by making measurements at $\theta=0^0$ and $\theta=\pi/2$, it would be possible to observe how the count rate would change. If only one boson type exists of given parity, the count rate should increase with $\theta=0^0$ for a pseudoscalar boson, and with $\theta=90^0$ for a scalar, and the count rate at the angle $\theta$ giving minimum signal should be compatible with noise.

Let us make the hypothesis of having verified that the rotation of the polarization vector in the two directions $\theta=0^0$ and $\theta=90^0$ respectivly maximizes and minimizes (or viceversa) the count rates with equal settings of the right and left reconversion beam lines, and that the minimum observed be compatible with zero. Having verified this hypothesis, one could choose the setting of the magnets in the conversion region that maximizes the signal rate and set the left string of reconversion magnets with the same setting as that of the conversion magnets.

Several measurements performed with this fixed setting of the conversion and left reconversion magnets and with increasing number of powered magnets in the right string of reconversion magnets would provide the experimental input for a number of equations of the type

$$\frac{S_R}{S_L} = \frac{\eta_R}{\eta_L} \sin^2(\frac{m^2 L_R}{4E}) / \sin^2(\frac{m^2 L_L}{4E}) \qquad (10)$$

where the only unknown is the mass m.

By plotting $S_R/S_L$ as a function of $L_R$, the m value which best fits the data could be determined and would not depend on the parameters W and N of the optics. Two measurements at least with different $L_R$ would be necessary to determine m.

By injecting the obtained m value into the relations (3) one would determine $g_{m\gamma\gamma}$. Notice that the error on the determination of $g_{m\gamma\gamma}$ is dependent on the accuracy of the knowledge of the parameters W and N of the optics.

If $L_L \leq L_R << 2\pi E/m^2$ the reconversion probabilities do not depend on m. Under these circumstances equation (10) reduces to

$$S_R / S_L = \eta_R L_R^2 / \eta_L L_L^2 \qquad (11)$$

and cannot be used to determine m, while equations (9) become

$$Ph_{L/R} = \frac{N}{32} \frac{W}{E} g_{m\gamma\gamma}^4 B^4 L_C^2 L_{L/R}^2 \qquad (12)$$

A single run with $L_L \neq L_R$ would be sufficient to verify whether equation (11) is satisfied. If yes, $g_{m\gamma\gamma}$ can be determined from equations (12).

A first test of the approach proposed in this paper could be performed in the PVLAS experiment by installing reconversion permanent magnets above and below the ellipsometer that embraces the superconducting conversion magnet. Space is available to install a 0.5 m long permanent magnet below the ellipsometer and up to 2 m long permanent magnets above the ellipsometer. The counting rate of photons regenerated above the ellipsometer could then exceed the counting rate of photons regenerated below the ellipsometer by a factor up to 16.

The expressions that give the probabilities of conversions and reconversions have been derived under the hypothesis that relativistic spin-zero bosons propagate in space following the Klein-Gordon equation of motion. Furthermore the fringe fields of the magnets have been ignored, and we have considered so far configurations of magnets in each string powered with magnetic field vectors all parallel.

By keeping the conversion region and the left reconversion string always powered in the same way, and by powering the magnets in the right string with various polarities it would be possible to check experimentally the laws of propagation of spin zero bosons in presence of magnetic fields [33,34] and to perform interference measurements using the magnets and their polarities in analogy to slits in interference experiments in optics.

For instance it will be interesting to check whether with only two magnets powered in the right string with opposite polarities the intensity of the regenerated beam would be zero, and to measure the intensity of the regenerated beam between two dipole magnets powered with opposite polarities, by inserting a semitransparent mirror in the photon beam path.

Notice that one could install a second identical normalization station downstream of the first one in order to measure directly possible errors or fluctuations of the normalization, since the same boson beam would traverse practically unperturbed both stations. Also one could insert a normalization station between the conversion region and the string of reconversion magnets used for measurements, instead of having the normalization station on the opposite side of the FP. One could even imagine a scenario with two different experiments located at the two opposite sides of the FP each one instrumented with a normalization station before the measurement string. Both experiments would be illuminated by the same arrangement of laser, FP and conversion magnets.

More generally one could imagine a facility with laser, FP and conversion magnets playing a role analogous to that of a particle accelerator and several experiments in series on the two boson beam lines, each with its normalization and measurement stations. If the laser energy E is increased, the size of the facility grows linearly with E. The quality of the single photon detection can improve with increasing E. However the construction and operation of the FP may become unpractical because of the need of a longer and longer optical bench for the two FP mirrors.

If the signals of PVLAS will be confirmed to be due to the existence of light bosons of masses of order $10^{-3}$ eV, it will be still worth exploring the lower mass region, because PVLAS and a regeneration experiment optimized for the mass region $10^{-3}$ eV would be unaffected by bosons with mass below $10^{-4}$ eV.

Regeneration experiments suggested to cover the mass region below $10^{-4}$ eV [10,11] require a linear extention for which a FP is hardly conceivable. However the use of two reconversion stations in series downstream of the conversion string of magnets (one station kept fixed for normalization, and the second with variable number of powered magnets used for measurements) remains interesting, because it avoids the need of determining the luminosity of the conversion station for the measurements of m.

## Acknowledgements

I have profited from countless discussions with Emilio Zavattini, from the dialectic ambience of the PVLAS collaboration and the stimulating atmosphere of the neutrino factory working groups.